# Garnet-free optical circulators monolithically integrated on spatially modified III-V quantum wells

Parinaz Aleahmad, Mercedeh Khajavikhan, Demetrios Christodoulides, and Patrick LiKamWa
CREOL, The College of Optics and Photonics, University of Central Florida, Orlando, Florida 32816, USA

**Optical circulators are indispensable components in photonic networks that are aimed to route information in a unidirectional way among their N-ports[1,2]. In general, these devices rely on magneto-optical garnets[3] with appreciable Verdet constants that are utilized in conjunction with other elements like permanent magnets, wave-plates, birefringent crystals and/or beam splitters. Consequently, these arrangements are typically bulky and hence not conducive to on-chip photonic integration[4-6]. Of interest would be to devise strategies through which miniaturized optical circulators can be monolithically fabricated on light-emitting semiconductor platforms by solely relying on physical properties that are indigenous to the material itself. By exploiting the interplay between non-Hermiticity and nonlinearity, here we demonstrate a new class of chip-scale circulators on spatially modified III-V quantum well systems. These garnet-free unidirectional structures are broadband (over 2.5 THz) at 1550 nm, effectively loss-free, color-preserving, and in proof-of-principle demonstrations have provided 23 dB isolation when used under pulsed-mode conditions at milliwatt (mW) power levels.**

On-chip photonic networks hold great promise for enabling next-generation high speed computation and communication systems[4-6]. It is currently envisioned that future integrated photonic networks will be capable of processing digital information at high data rates on a single monolithic platform by involving a multitude of optical components ranging from lasers to modulators, to routers, interconnects and detectors.[7] For such a network to properly function, it is imperative to engage components like optical circulators in order to route data traffic to designated destinations. Circulators are nonreciprocal devices[1] that are meant to direct signals in a unidirectional fashion between successive ports, as shown schematically in Fig. 1a. Typically, this response results from time-reversal symmetry breaking as afforded by the Faraday effect, when manifested in magneto-optical materials like garnets[3] or ferrites (in the microwave regime). Unfortunately, standard optoelectronic semiconductor materials lack magneto-optical properties and hence cannot be directly used in this capacity. One possible avenue to overcome this limitation is to integrate magneto-optical garnets on either silicon or III-V substrates[8-12]. While in the last few years there has been significant progress along these lines, such arrangements still require the presence of an external permanent magnet. To address this issue, a number of garnet-free isolator designs have been recently suggested and implemented on silicon, based on parametric and nonlinear methodologies[13-20]. So far however, these approaches are met with a number of challenges either due to large footprint requirements or bandwidth limitations- especially when used in conjunction with micro-resonator arrangements. These problems become further acute, in more complex settings like that of an optical circulator where directional transport between a number of ports is expected to occur. In light of these difficulties, no broadband, on-chip, magnet-free circulator has been reported, as of yet. Clearly, of importance will be to develop new tactics through which unidirectional circulators can be miniaturized and readily integrated on a single wafer by utilizing physical processes that are intrinsic to the material itself.

The prospect of non-magnetic circulators was first introduced by Tanaka et al in 1965 within the context of radio frequency circuits[21]. These ferrite-free devices make use of active integrated transistors that are by nature highly nonlinear elements. At this point the question naturally arises as to whether similar strategies can be adopted in the optical domain by exploiting the synergy between amplification, loss, and nonlinearity. In this study we demonstrate a compact, non-magnetic, active 4-port circulator, monolithically integrated on an InP-based quantum well wafer. The operating principle behind this non-Hermitian device is the unidirectional optical transport enabled by a sequence of amplifiers and decoupling nonlinear elements. Note that all the physical properties needed to realize this structure (appreciable gain/loss and high nonlinearities) are amply available in this material platform. What made this multi-functionality possible on the same wafer, is a recently developed intermixing technique that allows one to spatially fine tune the bandgap energies so as to simultaneously accommodate semiconductor optical amplifiers (SOAs) along with lossy defocusing nonlinear components and transparent waveguide channels[22]. The demonstrated circulator provides 23 dB isolation in a broadband fashion when used under pulsed mode conditions at telecom wavelengths (1550 nm).

Figure 1 illustrates how this 4-port optical circulator functions. Signal injected into port 1 is first amplified in the initial SOA section. It is then directed into a lossy and highly nonlinear decoupler[23] (ND) that in turn routes it to a second ND unit by perturbing the balance of the two identical waveguide elements involved. What facilitates this process is the large nonlinearity offered by the quantum wells that responds to low power levels. At this stage, the signal enters the second ND unit after experiencing the same amount of loss, intentionally provided so as to break the symmetry of the optical transport. As a result, light now crosses over into the designated port 2 and its intensity is finally restored to its original power level after passing through a final SOA stage. In this same manner, signals from port 2 are directed to 3, etc. In all cases the transport is unidirectional in a counter-clockwise sense (1→2→3→4→1) as needed for the circulator to function.

In order to accommodate these different functionalities on the same wafer it is imperative to locally modify the bandgap energies of the quantum wells[24-26]. More specifically, the SOA sections are expected to operate at a smaller bandgap than the other two regions so as to obtain appreciable optical amplification while at the same time the energy gap in the passive waveguide channels must be substantially larger to minimize transmission losses. Meanwhile, the energy gap in the ND domains has to be optimized with respect to the operating wavelength for attaining strong levels of defocusing nonlinearities. To simultaneously satisfy these somewhat conflicting requirements on the same wafer, a special post-growth bandgap fine-tuning process has been utilized[22]. The untreated regions (involving six $In_yGa_xAs_{1-x}P_{1-y}$ quantum wells embedded in five barriers) have a bandgap energy of 0.813 eV, corresponding to an operating wavelength of 1525 nm (Fig. 2a). To increase transparency in the passive waveguide sections, the bandgap is blue-shifted to 1400 nm (0.886 eV) through thermally enhanced Ga out-diffusion that is promoted by $SiO_2$ capping layers (Fig. 2b). At the same time, during this thermal treatment, Ga out-diffusion is discouraged in the ND segments using $SiN_x$ capping layers, thus keeping the gap unchanged at 1525 nm (Fig. 2b). On the other hand, the bandgap is red-shifted in the SOA regions in order to introduce considerable gain at 1550 nm. In our studies, this task was accomplished after developing a new intermixing process based on Si-rich $SiN_x$ capping layers that unlike other techniques reduce the bandgap (in this case down to 1550 nm or 0.8 eV) (Fig. 2b). The photoluminescence response of these three distinct regions is shown in Fig. 2c. Following this treatment, the loss in the passive waveguides is brought down to 3 $cm^{-1}$ while it is increased to 22 $cm^{-1}$ in the ND regions. The device map is then transferred to the wafer using photolithography and reactive ion etching (Fig. 2b). Scanning electron micrographs of the various parts involved in this device are shown in Fig. 2d. After removing the capping dielectric layers, the device is spin-coated with benzocyclobutene (cyclotene) for planarization purposes and Au p-contacts are deposited on the SOA sections. The sample is then polished down to 120 μm, in order to prepare it for backside n-contact metal deposition. Finally the wafer is cleaved and mounted for characterization. More details concerning the fabrication processes can be found in Supplementary Information, section 1.

Before implementing this 4-port circulator, all its components were first individually designed, fabricated and characterized. The absorption loss of the nonlinear decouplers was experimentally measured at 1550 nm by monitoring the contrast of the corresponding Fabry-Perot resonances. On the other hand, the nonlinearity associated with such loss values (a direct outcome of Kramers-Kronig relations) was obtained by employing a free-space Mach-Zehnder interferometer. These measurements were carried out by acoustically modulating the output of a tunable erbium-doped mode-locked laser (pulse duration: 50 ps, repetition: 26 MHz). Using this procedure, the effective Kerr-nonlinear coefficient was found to be ~$2\times10^{-11}$ cm$^2$/W- indicating that this system can operate at small power levels. This large nonlinearity results from the combined processes of free-carrier-plasma and band-filling effects as well as exciton saturation[27,28] (Fig. 3a). Based on these measured values, the ND sections were judiciously designed so as to introduce unidirectional transport. As previously mentioned, in the first ND segment, the amplified signal is expected to remain in the same waveguide channel by nonlinearly breaking the degeneracy in the propagation eigenvalues as shown in Fig. 3a. On the other hand, in the ND configuration that follows, the attenuated optical pulses are meant to behave linearly in order to cross over to port 2 (Fig. 1b). Simulations indicate (Fig. 3b) that these two distinct responses can be accommodated with the same arrangement provided that the ND segments are 950 $\mu$m long when the injected power levels differ by 10 dB. This variation in power levels is provided through a sequence of SOAs and loss elements (NDs, mirrors, waveguides, etc). Given that the strong nonlinearity involved is of the defocusing type, the single-mode ND waveguide channels are appropriately designed so as to avoid leakage to radiation modes while allowing for strong coupling. The ND sections were further optimized after characterizing a number of fabricated devices of various lengths and waveguide separations (Fig. 3c). The mirror sections in this circulator system (required for compactness) were etched deeply only in the immediate vicinity of the sharp bends and had a measured reflectivity of 95%. The required 11 dB gain from the 1.2 mm long SOAs was obtained at 200 mA of bias current, corresponding to a 20 cm$^{-1}$ linear gain, again experimentally extracted from Fabry-Perot resonances. Our measurements indicate that all the elements involved meet the specifications required for this integrated circulator to function.

The unidirectional optical transport in the integrated device (Fig. 1b) was characterized using a mode-locked (50 ps), tunable erbium-doped fiber laser (1535 nm - 1565 nm). The laser pulses (with an average power of 1 mW) were coupled into the input ports of this system using a 40X microscope objective lens. The fidelity of this circulator (designed to be counter-clockwise) can be described through its power scattering matrix ($\tilde{S}$) that relates the output at each port in terms of the input power vector ($\vec{P}_{out} = \tilde{S}\vec{P}_{in}$):

$$\tilde{S} = \begin{pmatrix} 0 & c_b & 0 & c_f \\ c_f & 0 & c_b & 0 \\ 0 & c_f & 0 & c_b \\ c_b & 0 & c_f & 0 \end{pmatrix} \qquad (1)$$

where the matrix elements $c_{ij}$ describe the power transfer from input port $j$ to output port $i$. Given the geometry of the specific arrangement considered here (Fig. 1b), the cross-talk coefficients are only finite for neighboring ports. In equation (1), $c_f$ represents the intended forward (anti-clockwise) transmission coefficient while $c_b$ accounts for parasitic coupling effects (between the injected port and its clockwise predecessor). In this respect, the isolation fidelity can be measured in dBs using $10\log_{10}(c_f/c_b)$, which requires $c_f \gg c_b$ for best performance. The $c_f$ coefficients were experimentally obtained by measuring the output from channel 2 when port 1 is excited while the $c_b$ constants were determined by monitoring the leakage output from channel 4. From these, the isolation ratio can be directly evaluated. Subsequently, these measurements were repeated for all the other channels, always yielding identical results within ±2%. Figure 4a, illustrates the output power from terminal 2 (green curve) and 4 (red curve) as a function of the current driving the SOA right after the input, when channel 1 is excited. The insets in Fig. 4a, depict near-filed intensity patterns at the output of port 2. Clearly, the isolation ratio associated with this circulator increases with signal amplification that is proportional to the injection current. This is consistent with our previous arguments concerning the need for a nonlinearity-induced breaking of the symmetry in the ND section that follows. Moreover, as the SOA gain increases, the leakage tends to significantly decrease - further boosting the isolation ratio. For the structure considered here, optimum performance was achieved at a driving current of 200 mA. In this case the isolation ratio was found to be 23 dB at a wavelength of 1554 nm. The bandwidth of this circulator was also characterized by varying the wavelength of the mode-locked laser source used. Figure 4b shows that the system can function in a broadband fashion given that its isolation exceeds 18 dB over a wavelength range of 20 nm. The observed reduction in isolation, at both ends of the spectrum in Fig. 4b, is attributed to the finite bandwidth of the SOAs as well as the detuning of the ND sections involved. Measurements performed in this system also indicate that amplified spontaneous emission (ASE) resulting from SOAs has negligible impact on the circulator itself - as the ND sections are nonlinearly insensitive to this low spectral density noise. In our arrangement, ASE was removed at the outputs using bandpass filters. In general, these filters can be directly integrated in this arrangement using low-Q ring resonators. Finally, the circulator proposed here is designed to operate under pulsed conditions in return-to-zero (RZ) formats. In principle, the entering high-speed data streams can be synchronized (through appropriate delays) so as to avoid bit collisions in the ND sections.

In summary, we have experimentally demonstrated a magnet-free, monolithically integrated 4-port optical unidirectional circulator operating at 1.55 $\mu$m. This was accomplished by establishing amplification, attenuation, and nonlinearities on the same InP-based quantum well wafer through spatially modifying the bandgap energy. This nonlinear active circulator operates in a broadband manner with isolation ratios exceeding 18 dB over 20 nm. The demonstrated device has a small footprint and is directly compatible with

III-V semiconductor compounds. Of future interest would be to explore the possibility of using high contrast waveguide elements in order to further miniaturize such circulator systems. In this regard, the strong confinement can offer a number of advantages such as the realization of compact waveguide bends and enhancement of nonlinearities in the ND sections. Using similar principles in optical topological arrangements could also be another fruitful direction in attaining unidirectional energy transport [29,30].

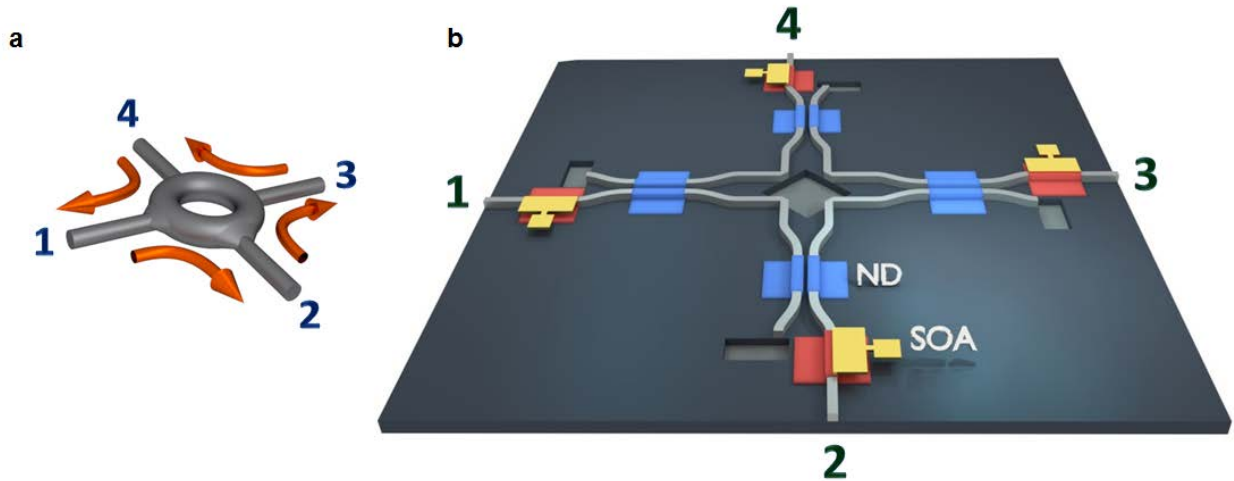

**Figure 1 | Optical circulator arrangements. a,** Conceptually, light is unidirectionally directed between ports in a circulator device. **b,** Schematic view of a magnet-free 4-port optical circulator used in this study. Signals in this active nonlinear system propagate in a counter-clockwise fashion. The gain (SOA) elements as well as the nonlinear decoupling (ND) segments are also depicted.

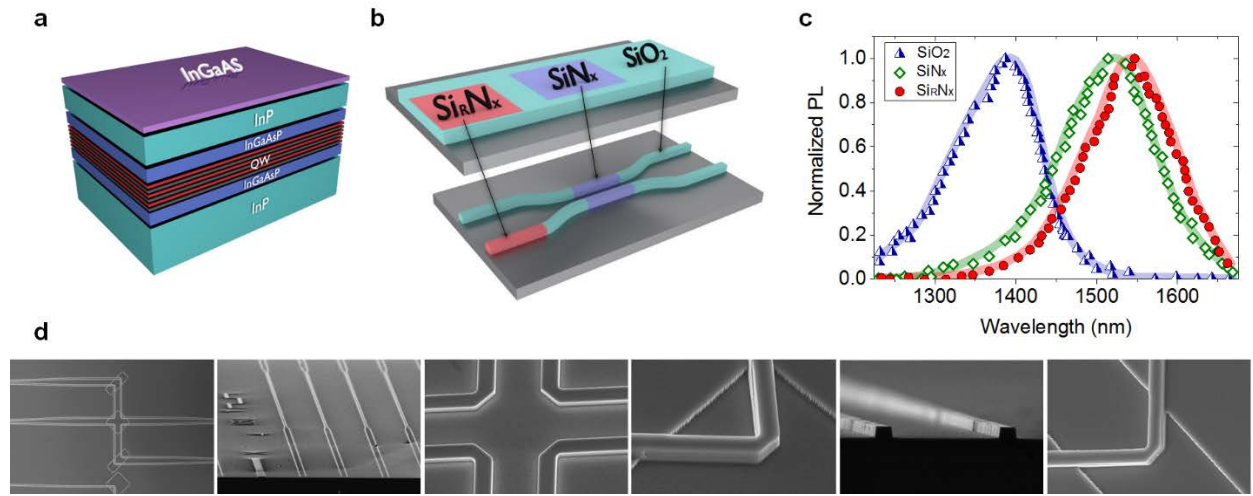

**Figure 2 | Fabrication process. a,** Cross-sectional view of the quantum well wafer used. **b,** Schematic illustration of the intermixing process. $SiO_2$, $SiN_x$ and Si-rich $Si_RN_x$ capping layers are employed to spatially fine-tune the optical bandgap in the different sections required for this circulator system. **c,** Photoluminescence response of the resulting three intermixed regions corresponding to ND (green), passive waveguide (blue) and SOA (red) segments. **d,** Scanning electron microscope pictures at various stages of the fabrication process. The different components needed for the 4-port circulator (SOAs, NDs, mirrors, etc) are depicted.

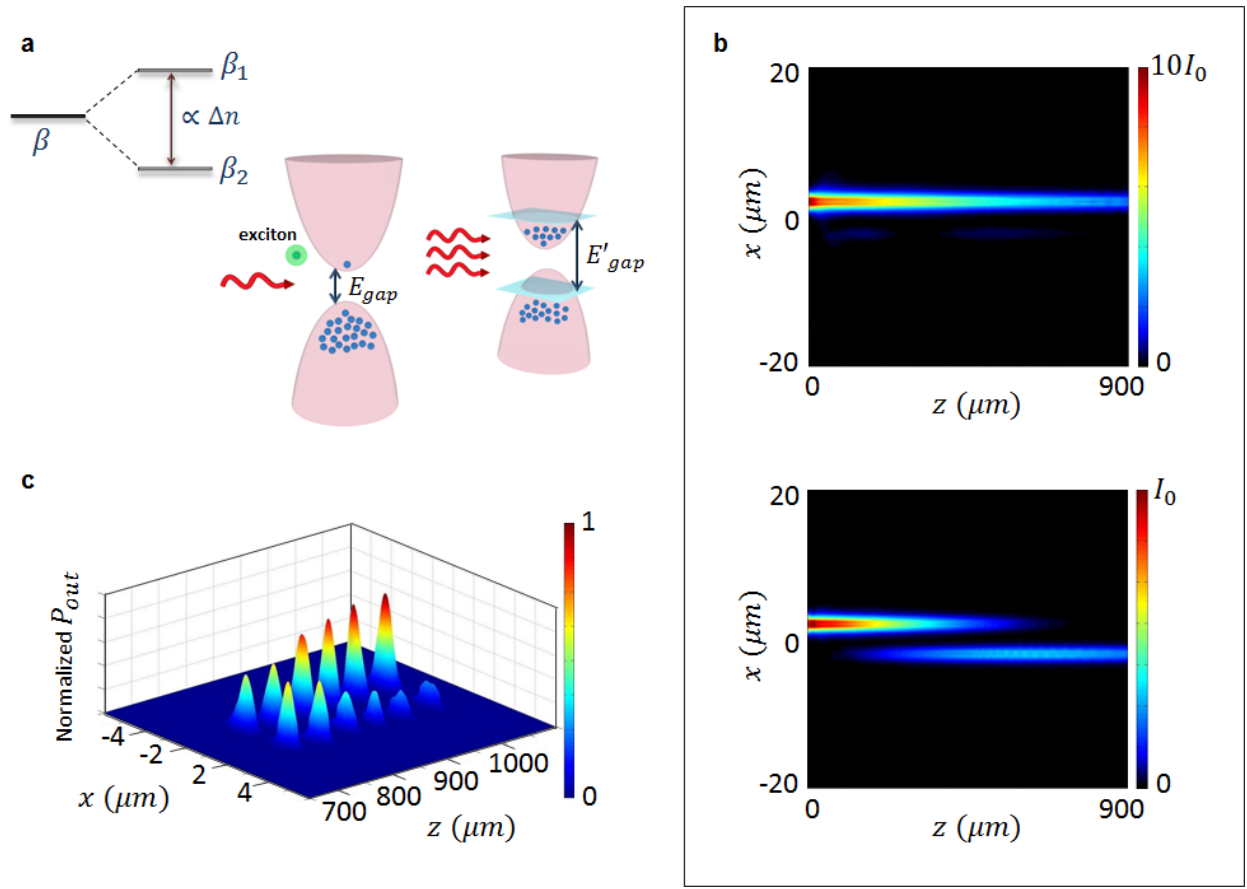

**Figure 3 | Optical propagation dynamics in the ND elements. a,** The large nonlinearity results from the presence of electrons in the conduction band because of absorption, leading to exciton saturation as well as free-carrier-plasma and band-filling effects. The degeneracy between the two propagation eigenvalues is nonlinearly broken. **b,** Simulations of optical transport in the ND elements under nonlinear (top) and linear (bottom) conditions. The amplified signal traverses the ND section within the same waveguide element because of nonlinear detuning (top), whereas at lower power levels the lightwave crosses over (bottom) in properly designed structures. **c,** Experimental measurements of the output power from ND arrangements having different lengths (linear regime). These measurements are used to fabricate and design a circulator with an optimum extinction ratio.

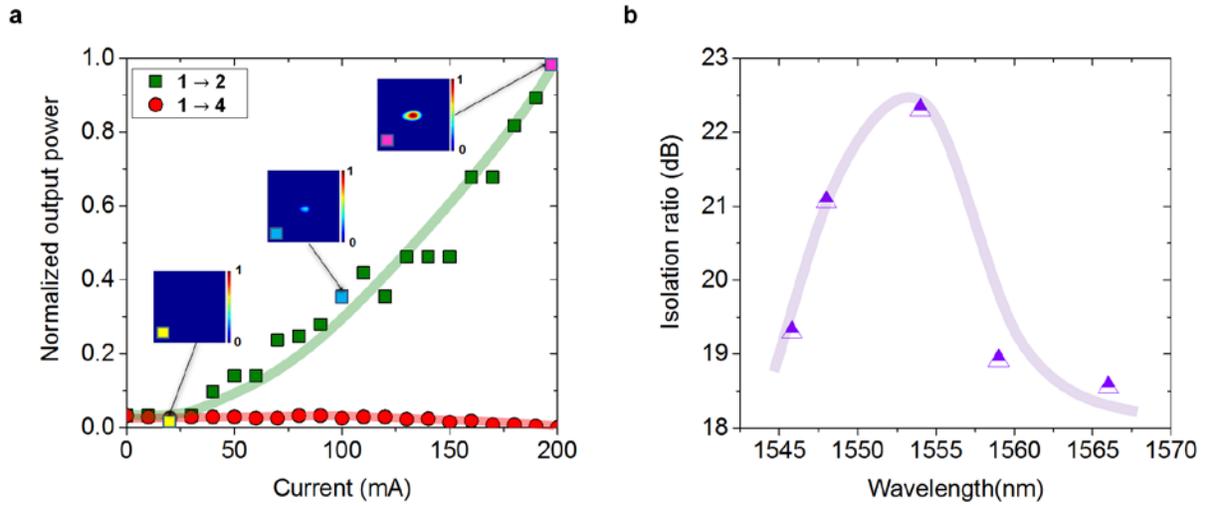

**Figure 4 | Experimental measurements. a,** Normalized output power from port 2 (green) and port 4 (red) as a function of the injected current in the SOA following port 1, where the signal is launched. The three insets show near-field intensity profiles from the output port 2, at different current levels. The isolation ratio is maximum at 200 mA. **b,** Wavelength dependence of the isolation offered by this 4-port circulator. The system operates in a broadband manner in the wavelength range of 1545-1565 nm.

## Acknowledgements

This work was partially supported by NSF (Grant No. ECCS-1128520) and AFOSR (Grant No. FA9550-14-1-0037).